\documentclass[]{spie}  

\usepackage{amsmath,amsfonts,amssymb}
\usepackage{graphicx}
\usepackage[colorlinks=true, allcolors=blue]{hyperref}

\usepackage{caption}
\usepackage{subcaption}

\usepackage{tikz,xcolor,hyperref}
\definecolor{lime}{HTML}{A6CE39}
\DeclareRobustCommand{\orcidicon}{%
    \begin{tikzpicture}
    \draw[lime, fill=lime] (0,0) 
    circle [radius=0.16] 
    node[white] {{\fontfamily{qag}\selectfont \tiny ID}};
    \draw[white, fill=white] (-0.0625,0.095) 
    circle [radius=0.007];
    \end{tikzpicture}
    \hspace{-2mm}
}
\newcommand{\orcid}[1]{\href{https://orcid.org/#1}{\orcidicon}}

\title{Laser-based metrology systems vs wavefront sensing techniques: a comparative overview between the Large Binocular Telescope and the Vera C. Rubin Observatory for the telescope alignment and collimation tracking}

\author[a]{Luca Rosignoli\orcid{0000-0002-0327-5929}}
\author[a]{Gabriele Rodeghiero\orcid{0000-0002-3469-9863}}
\author[b]{Sandrine J. Thomas\orcid{0000-0002-9121-3436}}
\author[c]{Guillem Megias Homar\orcid{0000-0001-6013-1131}}
\author[d,e]{Heejoo Choi}
\author[e]{John Hill}
\author[e]{Olga Kuhn\orcid{0000-0002-7389-1655}}
\author[k]{Elena Masciadri\orcid{0000-0002-0450-4092}}
\author[d,e]{Byeongjoon Jeong\orcid{0009-0000-8859-0575}}
\author[e]{Brandon Mechtley}
\author[e]{Christian Veillet\orcid{0000-0003-0272-0418}}
\author[q]{Elana Urbach\orcid{0000-0002-3205-2484}}
\author[r]{Brian Stalder\orcid{0000-0003-0973-4900}}
\author[e]{Jason Chu\orcid{0000-0003-3955-2470}}
\author[a]{Alessio Taranto\orcid{0009-0009-3271-3498}}
\author[g,s,u]{J. Bryce Kalmbach\orcid{0000-0002-6825-5283}}
\author[u]{Joshua E. Meyers\orcid{0000-0002-2308-4230}}
\author[g,h]{Andrew J. Connolly\orcid{0000-0001-5576-8189}}
\author[j]{Rebekah Polen\orcid{0009-0006-3437-9436}}
\author[s,t,u]{John Franklin Crenshaw\orcid{0000-0002-2495-3514}}
\author[g,h]{Krzysztof Suberlak\orcid{0000-0002-9589-1306}}
\author[b]{Tiago Ribeiro\orcid{0000-0002-0138-1365}}
\author[l]{Roberto Tighe}
\author[m]{Merlin Fisher-Levine}
\author[l]{Mario Rivera}
\author[n]{Enrico Giro\orcid{0000-0001-7301-8285}}
\author[o]{Rodolfo Canestrari\orcid{0000-0003-4591-7763}}
\author[l]{Holger Drass}
\author[l]{Pablo Zorzi}
\author[p]{Massimo Brescia\orcid{0000-0001-9506-5680}}

\affil[a]{\small INAF Osservatorio di Astrofisica e Scienza dello Spazio Bologna, Via P. Gobetti 93/3, 40129, Bologna, IT}
\affil[b]{\small Vera C.~Rubin Observatory Project Oﬀice, 950 N. Cherry Ave., Tucson, AZ 85719, USA}
\affil[c]{\small California Institute of Technology, 1200 East California Boulevard Pasadena, California 91125}
\affil[d]{\small Wyant College of Optical Sciences, University of Arizona, 1630 E University Blvd, Tucson, AZ 85721, USA}
\affil[e]{\small Large Binocular Telescope Observatory, University of Arizona, 933 N Cherry Ave, Tucson, AZ 85721, USA}
\affil[g]{\small DIRAC Institute, University of Washington, Seattle, WA 98195, USA}
\affil[h]{\small Department of Astronomy, University of Washington, Seattle, WA 98195, USA}
\affil[j]{\small Department of Physics, Duke University, Durham, NC 27708, USA}
\affil[k]{\small Observatorio Astrofisico di Arcetri, Largo Enrico Fermi 5, 50125 Florence, Italy.}
\affil[l]{\small Vera C. Rubin Observatory, Avenida Juan Cisternas 1500, La Serena, Chile}
\affil[m]{\small Department of Astrophysical Sciences, Princeton University, Princeton, NJ 08544, USA}
\affil[n]{\small INAF OATS, Via Giovan Battista Tiepolo 11, 34143, Trieste, Italy}
\affil[o]{\small INAF IASF, Via Ugo la Malfa 153, 90146, Palermo, Italy}
\affil[p]{\small Department of Physics ``E. Pancini”, University Federico II of Napoli, Via Cintia, 80126 Napoli, Italy}
\affil[q]{\small Harvard Physics Department, 17 Oxford St. Cambridge MA USA}
\affil[r]{\small Vera C Rubin Observatory/NOIRLab, Tucson, AZ}
\authorinfo{Further author information: (Send correspondence to L.R.)\\L.R.: E-mail: luca.rosignoli@inaf.it}
\affil[s]{\small Kavli Institute of Particle Astrophysics and Cosmology, Stanford, CA, 94305, USA}
\affil[t]{\small Department of Physics, Stanford University, Stanford, CA, 94305, USA}
\affil[u]{\small SLAC National Accelerator Laboratory, Menlo Park, California, 94025, USA}

\begin{document} 
\maketitle

\begin{abstract}
This work presents a comparative overview of the collimation and alignment strategies employed by two leading 8m-class facilities: the Large Binocular Telescope (LBT) and the Vera C. Rubin Observatory. While both telescopes share a challenging fast f-number of approximately f/1.2 (considering the LBT in its Prime Focus configuration), they have adopted reciprocal architectures for the initial optical alignment strategy and for maintaining collimation during the night.
As an initial alignment strategy, the LBT relies on a Wavefront Sensing technique called Focal Plane Image Analysis.
Conversely, the Vera C. Rubin Observatory baseline foresees the usage of a Laser Tracker system to establish the initial optical states.
The strategies for preserving the optical alignment and maintaining the collimation against gravitational flexure and thermal drift during observations are instead reversed. Besides the use of open-loop corrections based on Look-Up Tables, common on both telescopes, the LBT utilizes a laser-based Telescope Metrology System to monitor the relative position of optics in real-time, applying the corrections between the exposures. In contrast, the Rubin Observatory employs a Curvature Wavefront Sensing technique, using dedicated detectors at the four corners of the focal plane.
Rather than identifying a best strategy, this work aims to synthesize the strengths, limitations, and operational trade-offs of these complementary approaches, from the perspective of the next generation of Extremely Large Telescopes and their instruments.
\end{abstract}


\section{INTRODUCTION}\label{sec:intro}
The development of ground-based optical and infrared telescopes has undergone a paradigm shift with the advent of Active Optics Systems (AOS). Historically, telescopes relied on thick, rigid primary mirrors supported passively—an engineering approach that limited the maximum achievable diameter to approximately 5 meters due to prohibitive weight and resulting gravitational flexure \cite{wilson_active_1987}. The transition to the ``active" paradigm overturned this logic, enabling the use of thin, flexible mirrors whose shape and position are constantly maintained and optimized via a system of force actuators and hexapods. Modern AOS typically operate at low temporal frequencies ($f < 1$ Hz) to counteract static or quasi-static aberrations introduced by structural flexure under gravitational loads, thermal deformations, and mechanical drifts \cite{noetheActiveOpticsModern2001}. Today, the implementation of robust alignment strategies and active collimation control systems is a fundamental requirement to ensure that 8-to-10-meter class facilities consistently operate at their theoretical diffraction or seeing limits. 

This engineering necessity will reach even more critical proportions with the upcoming generation of 30-to-40-meter class Extremely Large Telescopes (ELTs) like the ESO ELT \cite{padovaniExtremelyLargeTelescope2023}, the Giant Magellan Telescope (GMT)\cite{johnsGiantMagellanTelescope2012a} and the Thirty-Meter Telescope (TMT) \cite{sandersThirtyMeterTelescope2013}. The order-of-magnitude increase in size and mass compared to current observatories will require a drastic escalation in the complexity of the electromechanical control systems needed to maintain opto-mechanical stability. These challenges will not be confined to the telescope structures alone but will inevitably extend to their complex focal plane instruments, whose physical volumes will be comparable to those of current 8-meter telescopes. A good example is MORFEO (Multiconjugate Adaptive Optics Relay for ELT Observations)\cite{magrinFinalOpticalDesign2024a}, a future ELT instrument designed to correct the atmospheric distortion over a large ($\sim1'$) Field of View (FoV). It will be a optical path of over 40 meter and a bench sizes of 8x7x3 meter. Its perfect and continuous opto-mechanical alignment will be vital to meet the stringent astrometric accuracy and Wavefront Error (WFE) requirements for first-light scientific instruments such as MICADO\cite{sturmMICADOFirstLight2024}. This kind of delicate instrument would require an active control of the optical state, like the current 8-meter class telescope.

In this context, this work presents a comparative analysis of the AOS architectures of two 8-meter class facilities: the Large Binocular Telescope (LBT)\cite{hillLargeBinocularTelescope2010} and the Vera C. Rubin Observatory (Rubin Telescope hereafter)\cite{ivezicLSSTScienceDrivers2019}. Both facilities adopt hybrid approaches, combining laser-based metrology and optical wavefront sensing techniques to drive the systems toward their best image quality. 
The primary objective of this analysis is not to identify a superior approach Such a determination would be intrinsically flawed and conceptually misleading due to the substantial differences in optical designs, scientific goals (flexibility and targeted observations for LBT versus wide-field surveys for Rubin), environmental conditions at their respective sites, and their different stages of operational maturity: over two decades of consolidated operations for LBT versus the final commissioning and early observations for Rubin. 
Rather, the goal is to highlight the optical performance, operational dynamics, and potential limitations associated with each architecture. By evaluating the balance between the use of absolute metrology and wavefront sensors, this study aims to establish a solid framework and extract fundamental operational engineering knowledge. These work serves as a technological bridge for the development, risk mitigation, and planning of future opto-mechanical strategies in the ELT era and for its cutting-edge instrumentation.

\subsection{LBT's AOS system}\label{subsec:lbt_aos}
The LBT, located on Mount Graham in Arizona, represents one of the most advanced optical/infrared facilities in the 8-meter class. Its unique binocular design features two 8.4-meter primary mirrors mounted on a single azimuthal structure, providing a combined collecting area equivalent to an 11.8-meter telescope and the resolving power of a 22.7-meter interferometer\cite{hillLargeBinocularTelescope2010}. To achieve and maintain its optical performance, LBT relies on a highly sophisticated AOS designed to continuously correct for static and quasi-static opto-mechanical aberrations caused by gravitational flexure, thermal variations, and mechanical drifts. 
Both primary and secondary mirrors are equipped with hexapods and axial actuators to move and shape the optics. In addition, the secondary mirrors, also called Adaptive Secondary Mirrors (ASMs) serve as both the secondary mirror and the wavefront corrector for the atmospheric turbulence, integrating AO capabilities directly into the telescope’s main optical train\cite{espositoLargeBinocularTelescope2011}.

The AOS of LBT comprises a unique sub-system in the field of astronomical telescopes: a laser-based metrology system that can provide real-time, high-precision monitoring of the relative positions of the telescope's optical components, also called Telescope Metrology System (TMS). This system is developed in partnership with the Giant Magellan Telescope Observatory (GMTO): at the LBT, the TMS is developed and is currently used so far between the M1 and the LBC of each side of the telescope \cite{rakichLasertrussBasedOptical2020,rakichCommissioningLaserMetrology2022,rodriguezImplementationLasertrussBased2020}, while at GMTO the TMS will be used between each primary mirror and its corresponding secondary mirror \cite{rakichPrototypingGMTTelescope2018,xinDevelopmentGMTTelescope2024}.
The TMS employs a multi-channel interferometer from Hexagon called Etalon Absolute Multiline Technology (EAMT). This
system is designed to measure absolute distances between fixed points, in particular between a set of collimators, which shoot laser beams, and reflective targets, usually a Spherically-Mounted Retro reflector (SMR). The EAMT operates at a specific wavelength band of 1532 +/-70 nm. It achieves remarkable precision measuring lengths of approximately 10 meters with a repeatability of 1-2 microns accuracy, and providing data flow at $\sim$10 Hz\footnote{For more details about the TMS at LBT, please refer to \citenum{jeongRealtimeCollimationCorrection2025}.}.
The information about the relative position between the optics coming from the TMS is used to calculate and send corrections to the hexapods in between the exposures.

Another important piece of the LBT AOS is the alignment strategy adopted for the LBCs, where it is not possible to use any dedicated wavefront sensor to map the aberrated wavefront. In this case, the determination of low-order aberrations is done using the science instruments themselves and a geometrical-based wavefront sensing method called Focal Plane Image Analysis (FPIA)\cite{wilsonReflectingTelescopeOptics1999}. This technique look at highly defocused stars (i.e. donuts, a proxy of the pseudo-pupil image) and can give an estimate of the aberrations at the focal plane by a quantitative measurement of their external and internal borders \cite{hillPrimeFocusActive2008}.

The paradigm adopted at LBT starts with the telescope alignment using the FPIA after which the TMS measures the relative position between the M1 and LBC and stores that as a reference. From thereon, TMS maintains that position, sending corrections for any appreciable drift in between the exposures. Before the adoption of the TMS the FPIA was needed roughly every 30 minutes, and after it has been pushed to 1.5-2 hours intervals. The synergy between the TMS and the FPIA allows LBT also to decouple the alignment-induced aberrations from those coming from other sources, like the spherical aberration due to differential temperature gradient across the M1 surface \cite{rakichCommissioningLaserMetrology2022}.

So far, the TMS has been implemented only between M1s and LBCs but a commissioning of a new configuration between the M1s and ASMs is ongoing, opening up the possibility of using the TMS with the Gregorian instruments\footnote{For more details about the TMS in M1-ASM configuration please refer to \citenum{jeongRealtimeCollimationCorrection2025,rosignoliModelingMetrologySystems2025}.}.

\subsection{Rubin's AOS system}\label{subsec:lsst_aos}
The Vera C. Rubin Observatory, located on Cerro Pachón in Chile, hosts the 8.4-meter Simonyi Survey Telescope, an engineering facility characterized by a three-mirror anastigmat optical design specifically conceived to conduct the Legacy Survey of Space and Time (LSST) \cite{ivezicLSSTScienceDrivers2019}. Unlike traditional observatories designed for targeted observations, the Rubin Telescope operates in survey mode, and it is optimized for an exceptionally wide Field of View (FoV) of 3.5 degrees. It employs an innovative monolithic primary-tertiary mirror (M1M3)\cite{tuellFabricationLSSTMonolithic2012} coupled with the largest convex secondary mirror (M2) ever built\cite{rodeghieroVeraRubinsM22024} and a 3.2 Gpix camera, the largest ever built as well\cite{gilmoreLSSTCameraOverview2008}. 
To maintain seeing-limited image quality across such a vast focal plane ($\sim60$ cm), the telescope requires a highly responsive AOS capable also of accommodating the survey's rapid pointing cadence of 30 seconds\cite{biancoOptimizationObservingCadence2021}.
The AOS manages the figures of the three mirrors (M1M3 + M2) thanks to axial actuators, and the collimation of the M2 and Camera hexapods w.r.t. M1M3 for a total of 50 degrees of freedom, sending the corrections in between the exposures\cite{neillOverviewLSSTActive2014, thomasRubinObservatorySimonyi2023b}.

Conversely to LBT, the paradigm adopted by Rubin aims to use a laser-based metrology system to achieve the initial coarse alignment, putting the telescope in the capture range of the AOS closed loop. The laser-based metrology system implemented is the Laser Tracker (LT), which can determine the 6 DoF pose of M2 and camera components relative to M1M3 with an uncertainty of approximately 40 $\mu$m and an angular precision of about 2 arcseconds \cite{homarIntegrationOpticalAlignment2024}.
To ensure that the AOS keeps the telescope at the seeing limit throughout the night, the Rubin's AOS closed loop is driven by a continuous stream of wavefront sensing measurements. The LSSTCam is equipped with four pairs of staggered ($\pm1$ mm) sensors w.r.t to the science plane array, located at the four corners. The technique adopted is the Curvature Wavefront Sensing (CWFS), based on the work of Roddier\citenum{roddierWavefrontReconstructionDefocused1993}. This approach derives phase information from the intensity variations from two or more images acquired in different focus positions. Even if it requires multiple images, it can describe the aberrated wavefront with a high number of modes\footnote{At Rubin is up to 28th Zernike modes.} and it offers some advantages over other wavefront sensing methods for wide-field survey telescopes\cite{xinCurvatureWavefrontSensing2015}. 
By relying on equally defocused intrafocal and extrafocal images, CWFS can use area sensors with relatively large fields of view. This allows significant flexibility in selecting reference sources to use for wavefront measurement when the source scenery is constantly changing from one visit to the next with the LSST cadence \cite{xinCurvatureWavefrontSensing2015}.
Moreover, the algorithm has been optimized to deal with some peculiar features of this telescope like the extremely fast f-number ($F/1.23$), a large central obscuration ($\sim 60\%$), and significant vignetting at the corner locations of the sensors and blended sources. For a comprehensive overview of the state-of-the-art CWFS technique developed for the Rubin Telescope, please refer to \citenum{xinCurvatureWavefrontSensing2015,janishAlgorithmRapidMeasurement2012,crenshawUsingAIWavefront2024,homarAdvancingVeraRubin2024,megiashomarActiveOpticsSystem2024}.

\section{THE DATASETS}\label{sec:dataset}
The analysis carried out by this work exploits a variety of different data, from the telemetry of the observatory subsystems to the science pipelines. In this section, we describe the types of data used and where they come from.

The data retrieved for LBT are a combination of three sources. The FWHM and ellipticity of the infocus LBCs images came from the IQ inspection carried out by Christian Veillet, astronomer and former director of LBTO. His analysis goes through all the LBCs' images from the first light up today, but we limited the dataset to the start of 2023, the starting point of the TMS in science operations. Among the many outcomes of his analysis, we used the catalogs of point-like source detections made with \texttt{SourceExtractor} that contain the position, FWHM, and ellipticity of each detection. We also gathered the wavefront sensing data from the logs of the FPIA, stored in the telescope computers. In these logs, all the data related to each FPIA sequence, like timestamps, Zernike sets, and filter are stored. Finally, we retrieved the observatory telemetry, like the telescope elevation, the DIMM seeing measurements, and the temperatures related to the mirror and the ambient air, from the Data Mining System (DMS), a database that contains all the telemetry of the observatory subsystems. Overall, we collected data for roughly 6000 LBC images over 94 nights and almost 800 FPIA runs.

The majority of the data used for the Rubin come from the Consolidated Database (ConsDB)\cite{lim_consolidated_2025}. This database aggregates a variety of data for each LSSTCam visit. The reader can think of consDB as an incredibly large image header, that contains also information from all the observatory subsystems.
From consDB, we retrieved in particular:
\begin{itemize}
    \item FWHM and ellipticity of the science detector. We calculate statistics\footnote{Median, minimum, and maximum values.} for the FWHM and ellipticity across the entire FoV, as well as for data restricted to the central 3$\times$3 detector raft, which covers approximately the same sky area as the LBC cameras ($\sim 26$ arcmin), and the on-axis region corresponding to the central detector ($\lesssim 8$ arcmin).
    \item The seeing measurements coming from the DIMM instrument, and another independent estimate coming from the CWFSs. This estimate is the best fit value of the blurring convolution kernel to the detected donuts on the CWFSs, this is why it is dubbed DONUT BLUR (DB hereafter)\cite{janishAlgorithmRapidMeasurement2012}.
    \item Temperature measurements of the in-dome air and outside.
    \item Telescope elevation and filter band.
\end{itemize}
In addition, the temperature telemetry for the primary mirror comes from the Engineering Facility Databse (EFD), where we collected and averaged the temperature of all the thermocouples attached to the mirror substrate.
We filtered this huge amount of data, restricting only to those visits that belong to science observations\footnote{In particular, the ones acquired with the so-called Feature-based Scheduler (FBS).}, and we limit the dataset to the last months of early observations (from Jan. 1st 2026), to exclude the early phases where the telescope was significantly away from the IQ requirements. 
To further clean up the dataset, we applied a filter for the Shapelets score. This metric comes from the Shapelets decomposition of the measured PFSs. This decomposition relies on a complete, orthonormal set of 2D basis functions constructed from Laguerre or Hermite polynomials weighted by a Gaussian. A linear combination of these functions can be used to model any image, in a similar way to Fourier or wavelet synthesis\footnote{Please for more detail and a complete list of literature on this topic, visit \url{https://astro.dur.ac.uk/~rjm/shapelets/}.} The Shapelets score is essentially the sum of power of all the non-atmospheric-related modes of the decomposition, and it is a good indicator of well defined and symmetric PSFs. After all the cleaning, the total number of visits remained at around 7.500.

\section{STATISTICAL COMPARISON}\label{sec:stat_comp}
The primary objective of this analysis is to evaluate the stability and response of these AOSs across varying operational and environmental conditions. To provide a comprehensive overview, the investigation is structured into four distinct parts, each of which illustrates the performance trends for the following variables:
\begin{itemize}
    \item Seeing conditions: examining how the systems respond to, or are affected by, different levels of external atmospheric turbulence.
    \item Elevation range: isolating and analyzing the impact of the residual gravity-induced opto-mechanical flexures on the telescope structures.
    \item Thermal gradients between the mirror and the surrounding air, as well as between the inside and outside air. These variables are crucial to understand the effects of local mirror and dome seeing, and the efficiency of thermal stabilization and ventilation systems.
    \item Time elapsed since the last active collimation: quantifying the temporal degradation of the optical alignment since the last dedicated collimation procedure.
\end{itemize}

To reliably quantify the image degradation due to the AOS residual errors, we need to identify a metric that is as free as possible from systematics. A first attempt was made by means of the Root Mean Square Error of the Wavefront Error (RMS WFE) using the Zernike coefficients available from the wavefront sensing technique of LBT (the FPIA) and Rubin (by the Wavefront Estimation Pipeline - WEP).
Even though this metric is a good indicator of the intrinsic image quality of the telescope, ithe major problem was the bad sampling for LBT. While the Rubin can track the evolving wavefront aberrations during the exposures thanks to the CWFSs, obtaining a Zernikes set after each visit, LBT retrieves the Wavefront state only when an FPIA is run, limiting the data points to less than 10 per night. Therefore, the RMS WFE is evenly sampled by Rubin, while it is poorly sampled by LBT, reducing the available data points to around 750 instead of the almost 6000 available LBCs images. Moreover, the LBT dataset could be biased because the Zernikes are always measured at the best moment for the telescope, right after a collimation.

For these reasons, we decided to move from the Zernike coefficients and try to use directly the infocus PSFs of the science images, exploiting the full available datasets.
The measured FWHM at the science focal plane can be considered as a summation of multiple contributors, following the description of Racine \cite{racineMIRRORDOMENATURAL1991}:
\begin{equation}\label{eq:racine}
    FWHM_{meas}^{5/3} = FWHM_{seeing}^{5/3} + FWHM_{optical}^{5/3} + \alpha_m^{5/3}\Delta T_m^2 + \alpha_d^{5/3}\Delta T_d^{2}\,,
\end{equation}
where $FWHM_{seeing}$ is the seeing contribution\footnote{Already scaled for zenith and wavelength of observation.}, $FWHM_{optical}$ is the contribution due to the optical state of the telescope, and $\Delta T_{m,d}$ are the components related to the mirror and dome seeing respectively, where $\Delta T_m \equiv T_{mirror} - T_{dome}$ and $\Delta T_d \equiv T_{dome} - T_{ambient}$.
We are interested in evaluating the impact of everything that is not related to the seeing in the image quality. In particular, the residual error of the AOS, that contains the $FWHM_{optical}$, but also the mirror and dome seeing components. In addition, we decide to not extrapolate the mirror and dome seeing contributions even if we have reliable temperature measurements, because we do not know the correct coefficients for $\Delta T_m$ and $\Delta T_d$ for this observatory, and try to find their values is beyond the scope of this work.
We defined the term $FWHM_{AOS}$ as the sum of these contributions, that can be estimated by subtracting the estimated seeing from the measured FWHM:
\begin{equation}\label{eq:iq}
    FWHM_{AOS} = [FWHM_{meas}^{5/3} - FWHM_{seeing}^{5/3}]^{3/5}\,.
\end{equation}
This term includes residual aberrations left by the AOS both in terms of optical alignments (i.e. gravity compensation) and mirrors figures (i.e. temperature compensation), but also uncorrectable perturbations like wind shakes or tracking errors of the mount. For the latter, we assume that for both telescopes such type of sources are not significant, at least statistically over the entire dataset. In addition, the $FWHM_{AOS}$ comprehends also the mirror and dome seeing components. To understand how much these factors affect the image quality, we evaluated the $FWHM_{AOS}$ against $\Delta_T$ and $\Delta_d$.

\subsection{Seeing contribution}\label{sec:seeing_calib}
The major contributor in the measured FWHM should be the atmospheric turbulence, therefore it is important to properly estimate the seeing in the science images. The seeing estimation is always a delicate aspect of an image quality analysis, and it can be estimated in many ways. The most used method is the Differential Imaging Motion Monitor (DIMM)\cite{sarazinESODifferentialImage1990a}. This method estimates the image degradation due to the atmosphere in terms of the Fried parameter $r_0$\cite{friedStatisticsGeometricRepresentation1965a}, by measuring the variance of the differential image motion in two small apertures, usually cut out in a single larger telescope pupil by a mask\cite{tokovininDifferentialImageMotion2002}.

\begin{figure}[htbp]
    \centering
    \includegraphics[width=0.8\linewidth]{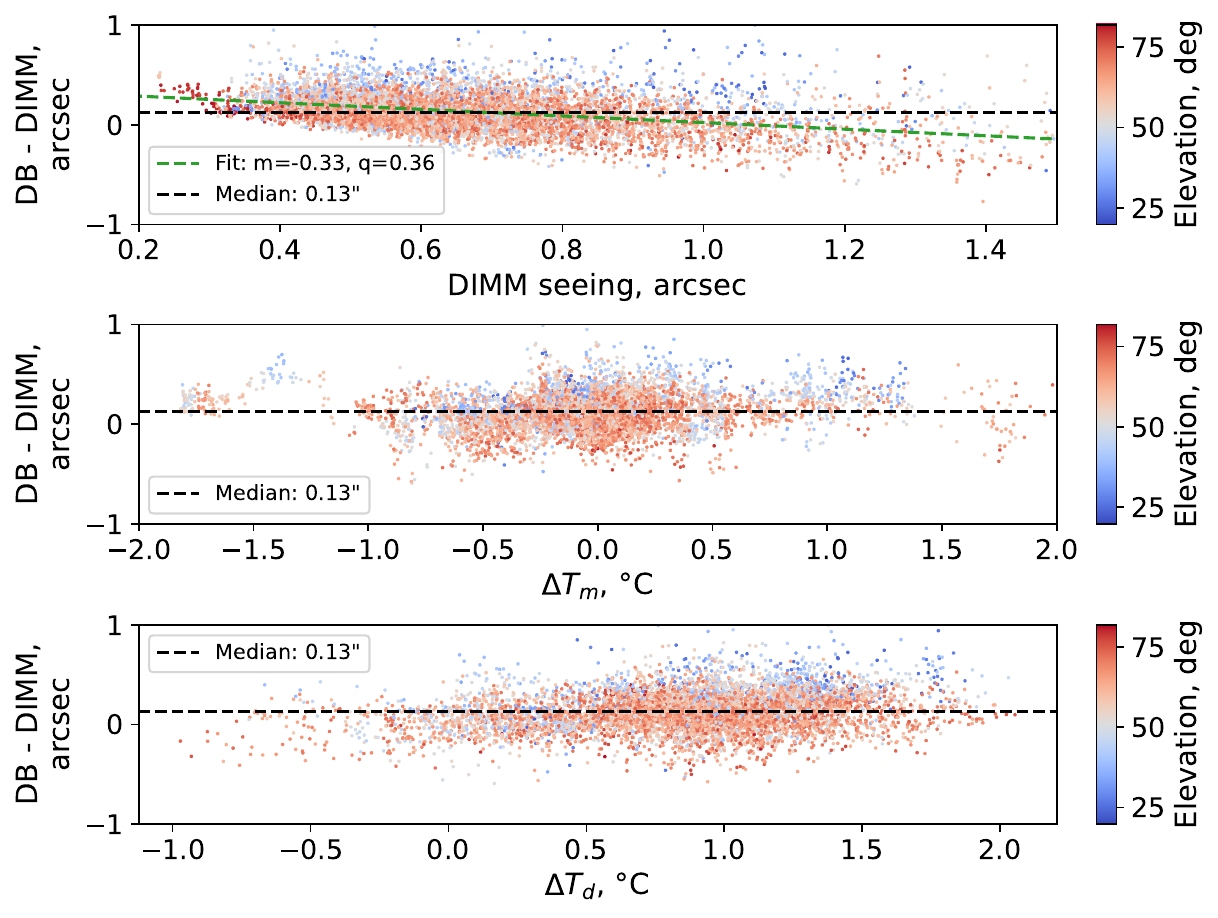}
    \caption{The difference between the Rubin's DB and the DIMM seeing estimates, against the DIMM itself, the $\Delta T_m$ and $\Delta T_d$ respectively. All the distributions are colored for the elevation.}
    \label{fig:seeing_diff}
\end{figure}

The variance of the differential image motion $\sigma_d^2$ (in square radians) is related to the Fried parameter $r_0$, the wavelength $\lambda$ for which this parameter is given, and to the sub-aperture
diameter D as
\begin{equation}
    \sigma_d=K\lambda^2r_0^{-\frac{5}{3}}D^{-\frac{1}{3}}\,,
\end{equation}
and the FWHM $\epsilon_0$ of the long-exposure seeing-limited PSF is computed with the standard formula:
\begin{equation}
    \epsilon_0=\frac{0.98\lambda}{D}=0.98\bigg(\frac{D}{\lambda}\bigg)^{0.2}\bigg(\frac{\sigma_d^2}{K}\bigg)^{0.6}\,.
\end{equation}
Usually, the DIMM seeing measurements $\epsilon_0$ are expressed at $\lambda=500$ nm and 1 air mass (a.m.). For this reason, it is not possible to directly subtract the DIMM seeing measurement as it is in equation \ref{eq:iq}, but it needs to be rescaled for the wavelength and a.m. of observation. Following the classic description of the atmospheric turbulence\cite{tatarskiWavePropagationTurbulent1961a,friedStatisticsGeometricRepresentation1965a}, the seeing $\epsilon_0$ depends on the air mass and the wavelength as:
\begin{align}
    \epsilon_0\propto a.m.^{0.6} \\ \epsilon_0\propto\lambda^{-0.2}\,.
\end{align}
If we limit the calibration to these two steps, usually the DIMM seeing is worse than the FWHM measured on the science images.
In the standard treatment of the DIMM, the Fried parameter $r_0$ is extracted following the Kolmogorov model\cite{sarazinESODifferentialImage1990a}, where the maximum scale length for the atmospheric turbulence is infinite. This assumption tends to overestimate the seeing contribution in the FWHM of the science detector for a large telescope, especially when the pupil diameter is a non-negligible fraction of the outer scale length\cite{tokovininDifferentialImageMotion2002,martinez_difference_2010}. Tokovinin 2002 \cite{tokovininDifferentialImageMotion2002} gives a first approximation of the seeing for a finite outer scale, following the von Karman model, and the conversion factor for the standard DIMM seeing is:
\begin{equation}\label{eq:tokovinin}
    \bigg(\frac{\epsilon_{vk}}{\epsilon_0}\bigg)^2 \approx 1 - 2.183 \bigg(\frac{r_0(\lambda)}{\mathcal{L}_0}\bigg)^{0.356}\,,
\end{equation}
where $r_0(\lambda)=\frac{0.98\lambda}{\epsilon_0'(\lambda,\,a.m.)}$ is the Fried parameter obtained from the DIMM seeing rescaled for the wavelength and a.m. of the science image, while $\epsilon_0$ is the DIMM seeing at 500 nm and 1 a.m.. Substituting in equation \ref{eq:tokovinin} we obtain:
\begin{equation}\label{eq:e0_to_evk}
    \epsilon_{vk} = \bigg[1 - 2.183\bigg(\frac{0.98\lambda}{\epsilon_0'(\lambda,\,a.m.)L_0}\bigg)^{0.356}\bigg]^{0.5}\epsilon_0\,.
\end{equation}
In this equation we need to assume a value for the outer scale $L_0$. The $L_0$ varies significantly over time and across different atmospheric layers, making it notoriously difficult to estimate precisely. Many works have been carried out to measure the outer scale, with different instruments and in different sites\cite{martin_first_1998,ziad_grating_2000,conan_wavefront_2002,ziad_comparison_2004} and the range of measured values span roughly between 20 to 70 meters. Consequently, adopting a unique value for $L_0$ represents a strong, albeit necessary, approximation for this analysis. 
Instead of arbitrarily chose a value, we investigate the distribution of the outer scales extracted directly from the datasets. Reverting Eq. \ref{eq:e0_to_evk} for $L_0$: 
\begin{equation}
    L_0 = \frac{0.98\lambda}{\epsilon_0'(\lambda,\,a.m.)}\bigg[\frac{1}{2.183}\bigg(1-\frac{\epsilon_{vk}^2}{\epsilon_0^2}\bigg)\bigg]^{-\frac{1}{0.356}}\,,
\end{equation}
and using the measured FWHM on the science images as $\epsilon_{vk}$, we obtained the outer scale distributions for both LBT and Rubin. In addition, we made the same estimates also for the Very Large Telescope (VLT) observatory, that we consider here as a reference, since it is a well characterize site in terms of environmental conditions and AOS performance\cite{guisardPerformanceActiveOptics2000a,guisardPerformanceImprovementActive2003,stephanLongtermPerformanceVLT2016}. We used the instrument-dedicated Image Quality database for FORS2 installed at the UT1, using data from Jan 2005 to October 2011.    
\begin{figure}[htbp]
    \centering
    \includegraphics[width=0.8\linewidth]{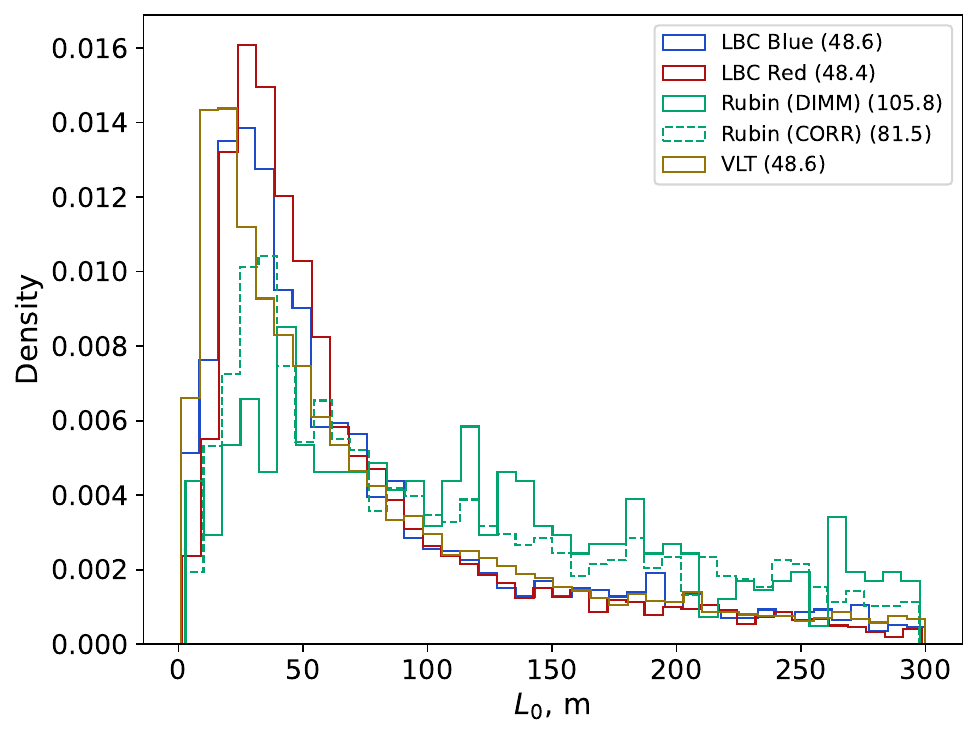}
    \caption{Figure showing the $L_0$ distribution for LBT (LBC - Red and Blue separately), Rubin and VLT (FORS2 on UT1). For Rubin also the outer scale distribution obtained with the corrected FWHM (CORR) is shown. The correction is the intercept value obtained by the linear regression between the DIMM seeing and the DB. For more details see Section \ref{sec:seeing_calib} (and Figure \ref{fig:seeing_diff}).}
    \label{fig:outer_scales}
\end{figure}
Figure \ref{fig:outer_scales} shows the outer scale distributions obtained with this method. The distributions of the two LBC channels are consistent to each other -- as expected -- as well as with VLT, with a median value around 48.5 meters, that falls inside the typical range of estimated outer scale in literature. The Rubin distribution instead, is more broad with a higher median value, around 105 meters. We suspect that the difference between Rubin and the other observatories could be due to a residual systematic in the measured FWHM of the Rubin's science images. We found evidences of such possible systematic among the results of this analysis and we tried to estimate it, converging to a value of 0.36 arcsec\footnote{Please see the discussion below in this section, in particular Figure \ref{fig:seeing_diff}}. We then try to calculate again the outer scale distribution of Rubin, by using the measured FWHM corrected\footnote{$FWHM_{corr} = [FWHM_{meas}^{5/3} - 0.36^{5/3}]^{3/5}$.} for this systematic. This new distribution, the Rubin (CORR) distribution in Figure \ref{fig:outer_scales}, moves slightly closer to the LBT and VLT ones, but still significantly higher. This supports the idea that the systematic resides on the measured FWHM but we are not totally accounting for it, and further investigation would be required. 
We then decided to not include the Rubin distribution, and use the average of the median $L_0$ values of LBCs and VLT ($L_0=48.5$ m) in the equation \ref{eq:e0_to_evk} to make the final calibration step, compensating for the non-infinite outer scale length, for all the observatories.
\begin{figure}[htbp]
     \centering
     \begin{subfigure}[b]{0.85\textwidth}
         \centering
         \includegraphics[width=\textwidth]{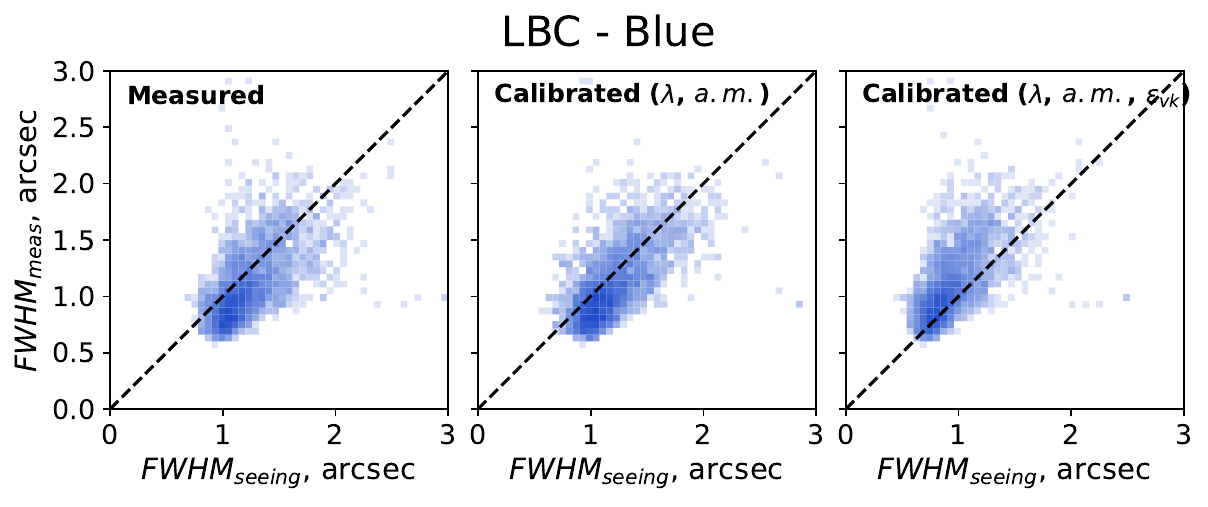}
         \caption{}
         \label{fig:lbcb_seeing_calib}
     \end{subfigure}
     \hfill
     \begin{subfigure}[b]{0.85\textwidth}
         \centering
         \includegraphics[width=\textwidth]{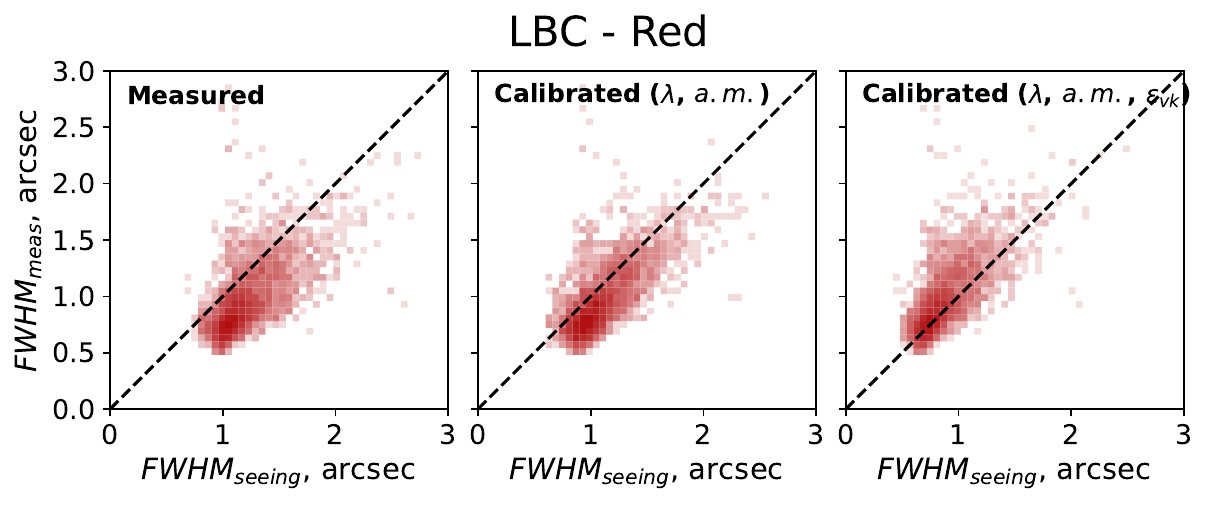}
         \caption{}
         \label{fig:lbcr_seeing_calib}
     \end{subfigure}
     \hfill
     \begin{subfigure}[b]{0.85\textwidth}
         \centering
         \includegraphics[width=\textwidth]{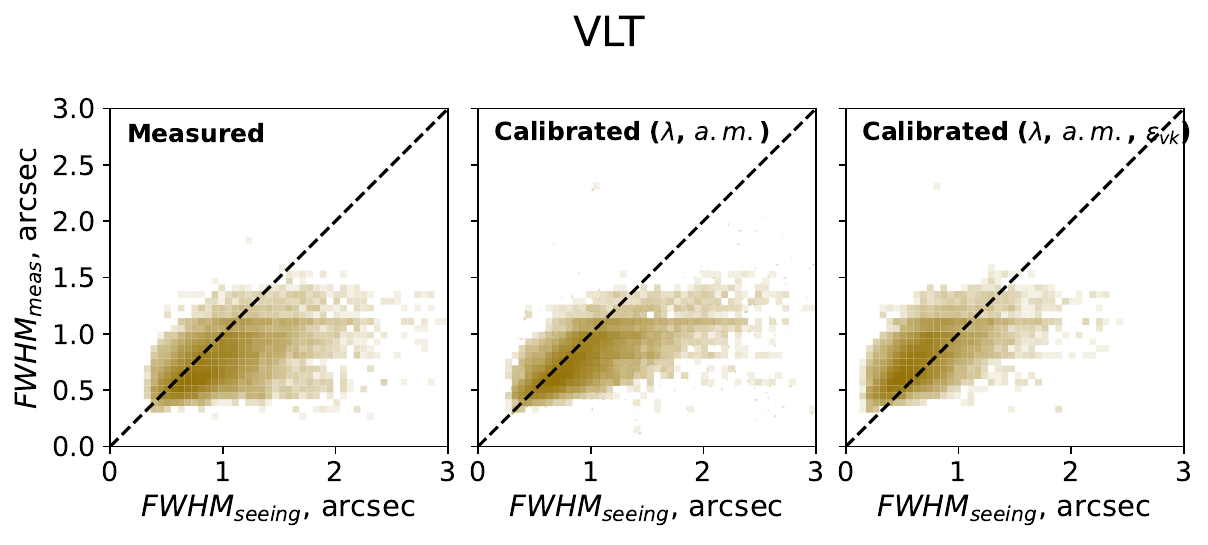}
         \caption{}
         \label{fig:vlt_seeing_calib}
     \end{subfigure}
\end{figure}
\begin{figure}[htbp]
    \centering
    \ContinuedFloat
     \begin{subfigure}[b]{0.75\textwidth}
         \centering
         \includegraphics[width=\textwidth]{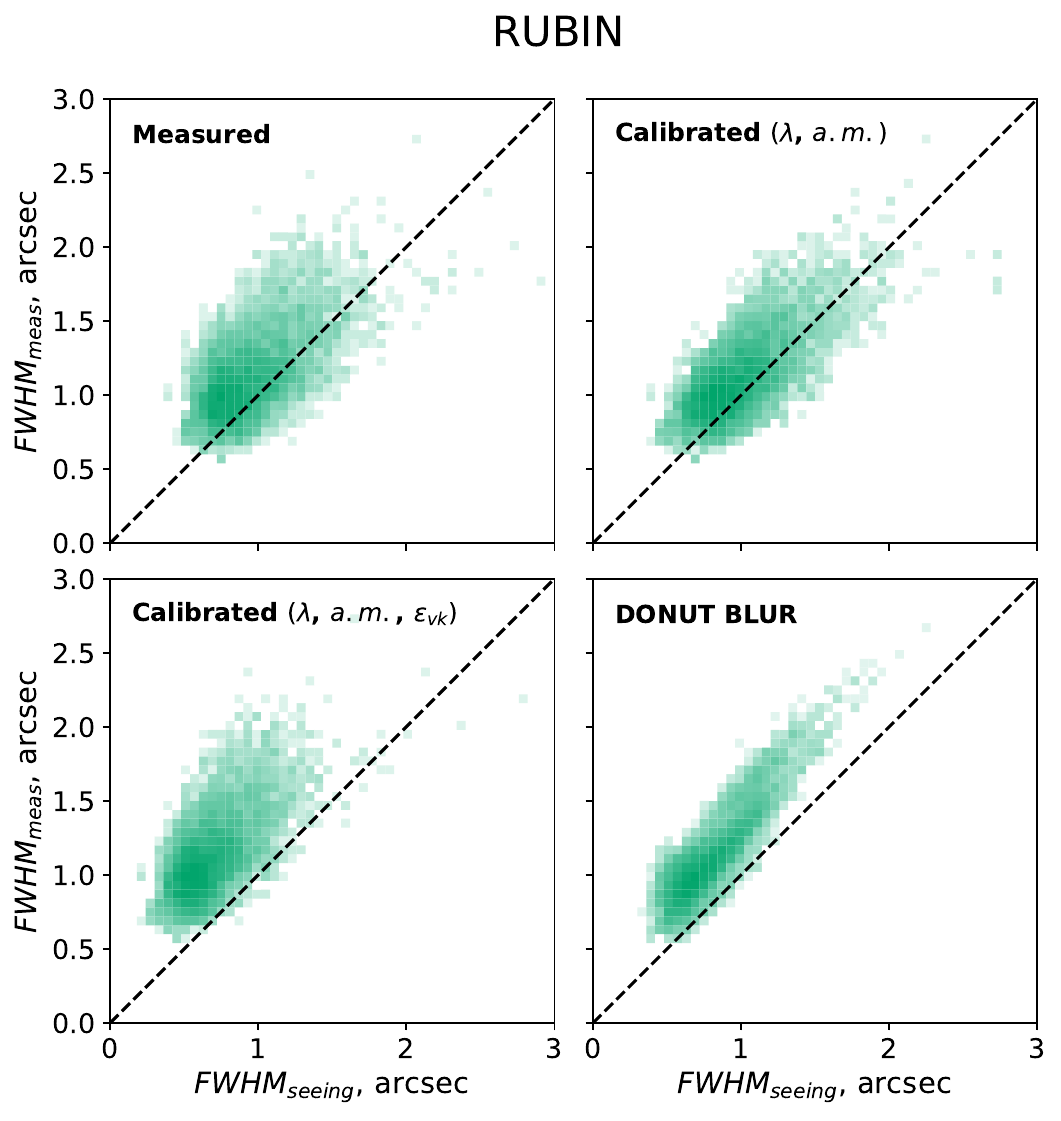}
         \caption{}
         \label{fig:lsst_seeing_calib}
     \end{subfigure}
        \caption{Comparison between the measured DIMM seeing (left), the calibrated one for $\lambda$ and a.m. (center), and calibrated also for the outer scale (right). The instrument order is LBC-Blue (a), LBC-Red (b), FORS2 (c) and LSSTCam (d). For LSSTCam, the comparison using the DB is also presented. The plots are 2D histograms, showing the distribution of the seeing value against the measured FWHM at the science detector (on-axis); the dashed line represents the 1:1 relation as a reference.}
        \label{fig:seeing_calib}
\end{figure}
Figure \ref{fig:seeing_calib} shows the different seeing estimates for LBC-Blue, LBC-Red, FORS2 and LSSTCam respectively. 
These results show the aforementioned behavior -- for LBCs and FORS2 -- in which the measured FWHM is statistically better than the seeing measured from the DIMM (Figure \ref{fig:seeing_calib} left plots). The calibrations for the wavelength and the a.m. have the effect to reduce the spread of the distributions (Figure \ref{fig:seeing_calib} center plots), while the correction for the outer scale have the significant effect of reducing the measured seeing, bringing the distributions closer to the identity (Figure \ref{fig:seeing_calib} right plots). 
It is important to stress again the critical adoption of a single value for the outer scale, that significantly impact the final seeing estimates. Therefore, the reader has to take these results with caution. They are not meant to estimate, in terms of absolute value, the performance of the AOS of LBT and Rubin.

The results shown in Figure \ref{fig:seeing_calib} support the idea that some systematic could be present in the Rubin images, where the $FWHM_{meas}$ is usually larger than the DIMM seeing, and this discrepancy is boosted with the calibration for the outer scale. 
Moreover, a significant difference is also present between the two Rubin seeing estimates (DIMM and DB).
In Figure \ref{fig:seeing_diff} the difference between the DB and the DIMM is evaluated against the DIMM seeing itself, the $\Delta T_m$ and the $\Delta T_d$, and also the telescope elevation.
These results show a median excess of the DB w.r.t. the DIMM of about 0.13 arcsec. Moreover, the difference converges to a value of $\sim$ 0.36 arcsec towards good seeing conditions. A slight dependence with the $\Delta T_m$ it is also visible, in particular for $\Delta T_m > 1.0$, suggesting a possible contribution of the mirror seeing effect as described in \citenum{racineMIRRORDOMENATURAL1991}, while no correlation with the $\Delta T_d$ looks appreciable. Finally, also a slight dependency with the elevation is visible, having larger values towards lower elevations.

We interpreted the difference between the DIMM and the DB as the cumulative effect of multiple factors:
\begin{itemize}
    \item \textit{Dome seeing + Mirror seeing}: A possible term of internal seeing, due to air turbulence inside the dome (dome seeing) and the not perfect thermalization of the primary-tertiary mirror (mirror seeing). The enclosure of the Rubin observatory is equipped with 34 louvers that are opened at the beginning of the night to enhance the thermalization of the telescope and permit a laminar flow above the optics. Since Oct. 2025 six of them were installed and other six became operational in May 2026. This reduced venting capability could induce unwanted dome seeing.
    \item \textit{Non-optimal AOS}: The not fully optimized AOS, including the open-loop static corrections -- also known as Look-Up Tables (LUTs) -- to compensate the gravity induced deflections and the thermal induced surface deformations of the opto-mechanical structure system.
\end{itemize}
These are symptoms related to the current stage of the Rubin observatory, that is facing the last part of the commissioning. For this reason it will be crucial to redo this work, once Rubin has at least one year of consolidated science operations.

\subsection{$FWHM_{AOS}$ metric}\label{sec:aos_fwhm}
\begin{figure}[htbp]
    \centering
    \includegraphics[width=0.7\textwidth]{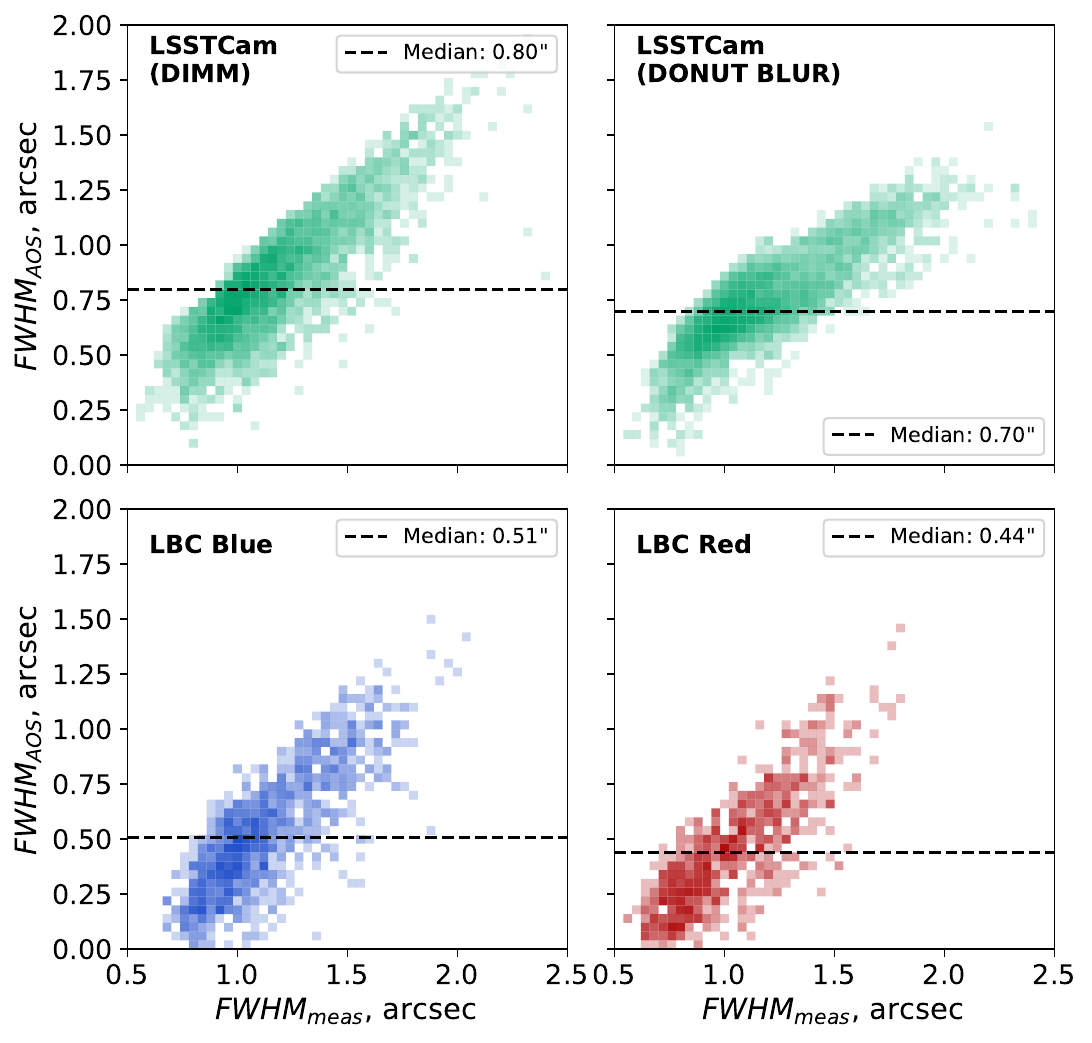}
    \caption{Figure showing the $FWHM_{AOS}$-$FWHM_{measured}$ relation for LSSTCam, LBC-Blue and LBC-Red respectively.}
    \label{fig:fwhm_aos_vs_measured}
\end{figure}

Having defined the $FWHM_{AOS}$ in equation \ref{eq:iq}, we inspected its contribution in the overall image quality budget. The dominant term in science PSF is the atmospheric seeing when the telescope is kept aligned by the AOS. In this seeing-dominated regime, the $FWHM_{AOS}$ is low and does not shows a significant correlation with the $FWHM_{meas}$. On the other side, if the optical state of the telescope start to diverge, the $FWHM_{AOS}$ became the dominant term in the PSF size, we can call this the AOS-dominated regime. In this regime the $FWHM_{AOS}$ and the $FWHM_{meas}$ are correlated.
In Figure \ref{fig:fwhm_aos_vs_measured}, we plot the $FWHM_{AOS}$ against the $FWHM_{meas}$.
In this Figure, the two regimes are merged with a smooth transition between them. At lower values of $FWHM_{AOS}$ the linear relation is lost and it begins to become gradually dominant for higher values. Moreover, the $FWHM_{AOS}$ obtained with the DB shows a non-linear trend in the AOS-dominated regime, highlighting its different nature from a pure seeing estimation method like the DIMM.

\subsection{Seeing dependency}\label{subsec:seeing_dep}
\begin{figure}[htbp]
    \centering
    \includegraphics[width=0.7\textwidth]{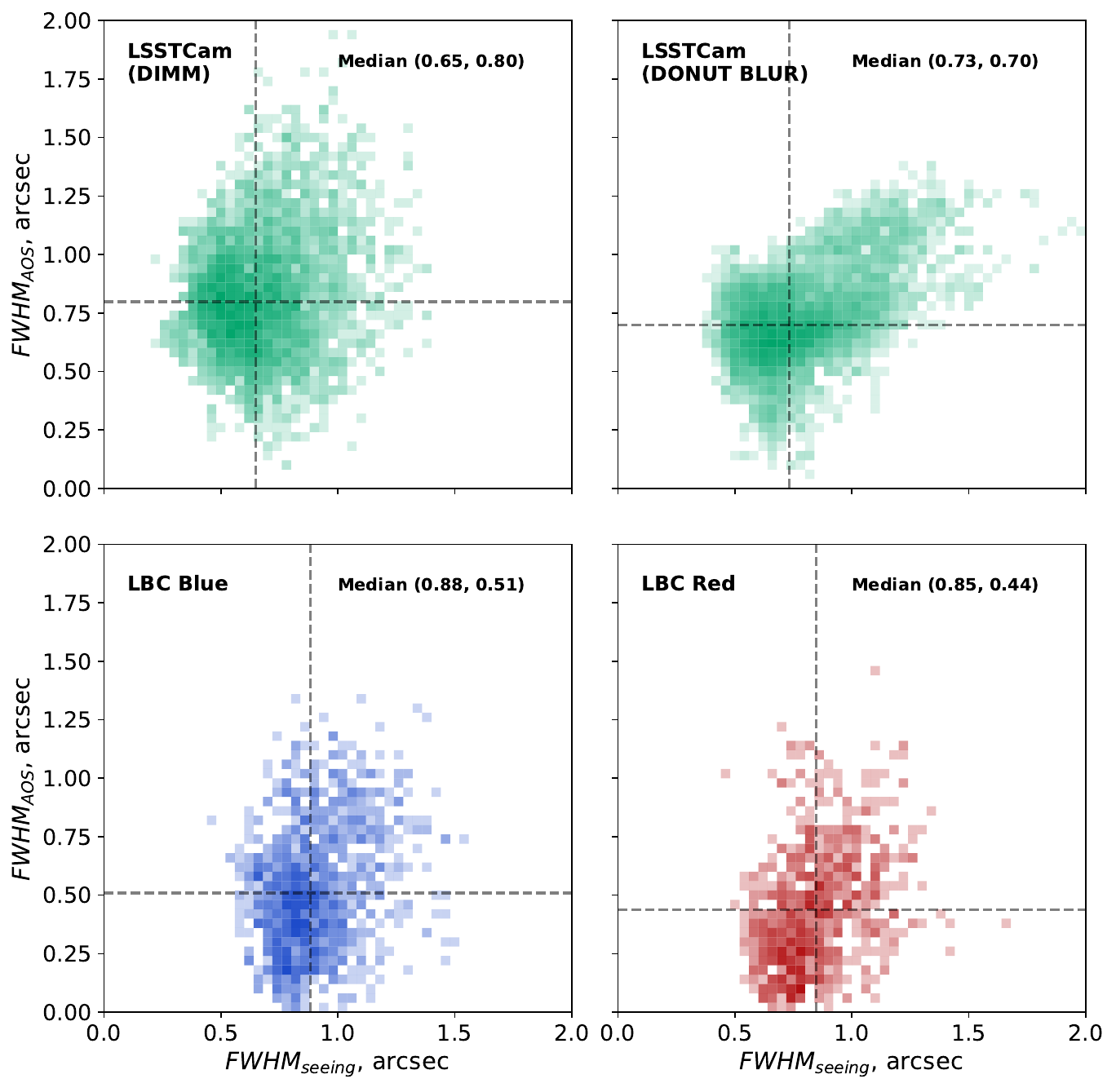}
    \caption{The relation between the $FWHM_{AOS}$ and the seeing ($FWHM_{seeing}$), for LSSTCam using the calibrated DIMM (top-left), LSSTCam using the DB (top-right), LBC-Blue (bottom-left) and LBC-Red (bottom-right) respectively. The position of the medians along the two axes are shown as reference.}
    \label{fig:fwhm_aos_vs_seeing}
\end{figure}

We evaluated the $FWHM_{AOS}$ contribution against the seeing measurement itself (Figure \ref{fig:fwhm_aos_vs_seeing}). This study should inspect the stability of the AOS systems in different seeing conditions. This is particularly critical for the Rubin AOS since its closed loop relies on the CWFSs. The LBT instead should be quite stable in this case, because its closed loop exploits the metrology of the TMS, even if it is limited only to the rigid body alignment of the optics. 
Figure \ref{fig:fwhm_aos_vs_seeing} shows no appreciable correlations in all cases, suggesting that both AOS systems are solid against the seeing conditions. The position of the median in both axes is taken as a reference. The median values in the x-axis ($FWHM_{seeing}$) are consistent with the literature values for both LBT\cite{turchiCharacterizationLBTAtmospheric2022} ($\sim0.85$ arcsec) and Rubin ($\sim0.7$ arcsec).
A non-negligible difference between the median positions for the two Rubin distributions (DIMM and DB) is present and it is close to the 0.13 arcsec highlighted in Section \ref{sec:seeing_calib}. This offset supports the idea that the DIMM is not sensitive to some contribution in the measured FWHM of the science images, while the DB is. Since the DB is an estimate of the blurring kernel applied to the modeled donuts\cite{janishAlgorithmRapidMeasurement2012}, the consistently higher DB w.r.t. the seeing supports the idea that the dome/mirror seeing as the probable contributor in the Rubin FWHM. The Rubin DIMM is located outside the main dome, therefore is not sensitive to the in-dome turbulence nor to the mirror seeing.

\subsection{Elevation dependency}\label{subsec:elevation_dep}
\begin{figure}[htbp]
    \centering
    \includegraphics[width=0.7\textwidth]{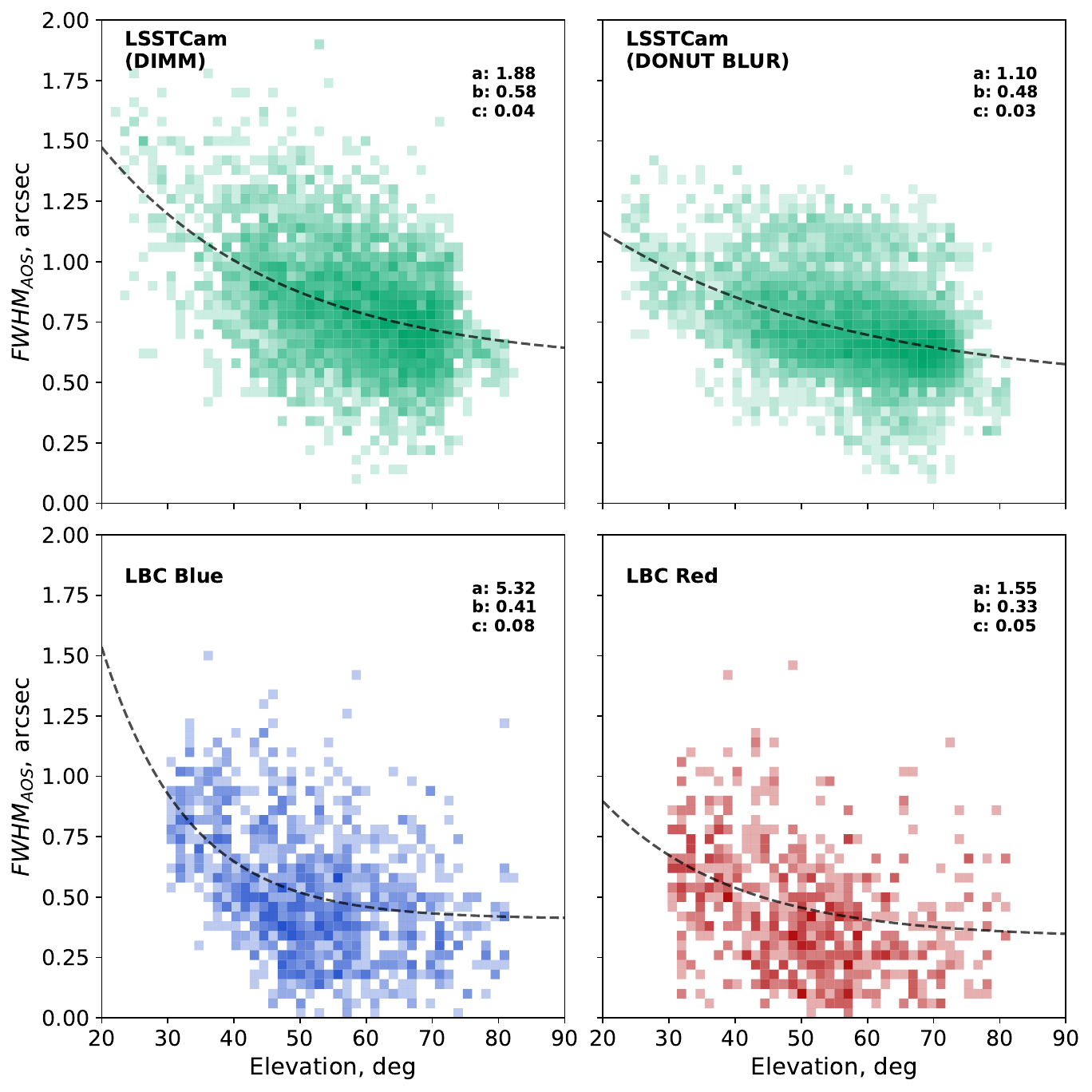}
    \caption{Relation between the $FWHM_{AOS}$ and the telescope elevation, for LSSTCam - DIMM (top-left), LSSTCam - DB (top-right), LBC-Blue (bottom-left), and LBC-Red (bottom-right), respectively. The exponential fit is shown and the best fit parameters are reported for each distribution.}
    \label{fig:fwhm_aos_vs_elevation}
\end{figure}

Evaluating the $FWHM_{AOS}$ along the elevation is fundamental to understand the ability of the AOS to keep the telescope aligned against the gravity load. With this analysis, we can evaluate both the correctness of the open-loop corrections and also the AOS closed loop capability to compensate for any residual errors. Figure \ref{fig:fwhm_aos_vs_elevation} shows a similar behavior for both Rubin and LBT. In any case, the $FWHM_{AOS}$ tends to increase at lower elevation and it stabilizes toward the zenith. We also tried to model the trends, giving a general function for the AOS response.
Since we did not have any prior information about the functional form of the fit, we evaluate the Bayesian Information Criterion (BIC)\cite{schwarz1978estimating} to understand the correct model to use avoiding overfitting. The exponential form resulted to be the best choice, therefore we modeled the trends using an exponential function:
\begin{equation}
    y = ae^{-cx} + b\,.
\end{equation}
In this case the systematic between the Rubin DIMM and the DB is not significant, suggesting that the contribution does not depend on the telescope elevation.
The plots of Figure \ref{fig:fwhm_aos_vs_elevation} may highlight also the benefit of the TMS for LBT. Since the laser-based metrology of the TMS is completely unaffected by the seeing conditions -- that degrades towards the horizon -- the LBT AOS close loop should perform better at lower elevation. This can be seen in the figure by the steeper shape of the exponential fit for LBCs, that keeps the $FWHM_{AOS}$ low for lower elevations, while for Rubin starts to increase earlier. It is also worth mentioning that the elevation range of LBT is smaller than the one of Rubin, and the absence of points at elevation lower than 30 deg could affect the exponential fit. Therefore this claim needs to be taken as only probable and requires further studies to be conclusive.

\subsection{Temperature dependencies}\label{subsec:delta_temp_dep}
\begin{figure}[htbp]
    \centering
    \includegraphics[width=0.75\textwidth]{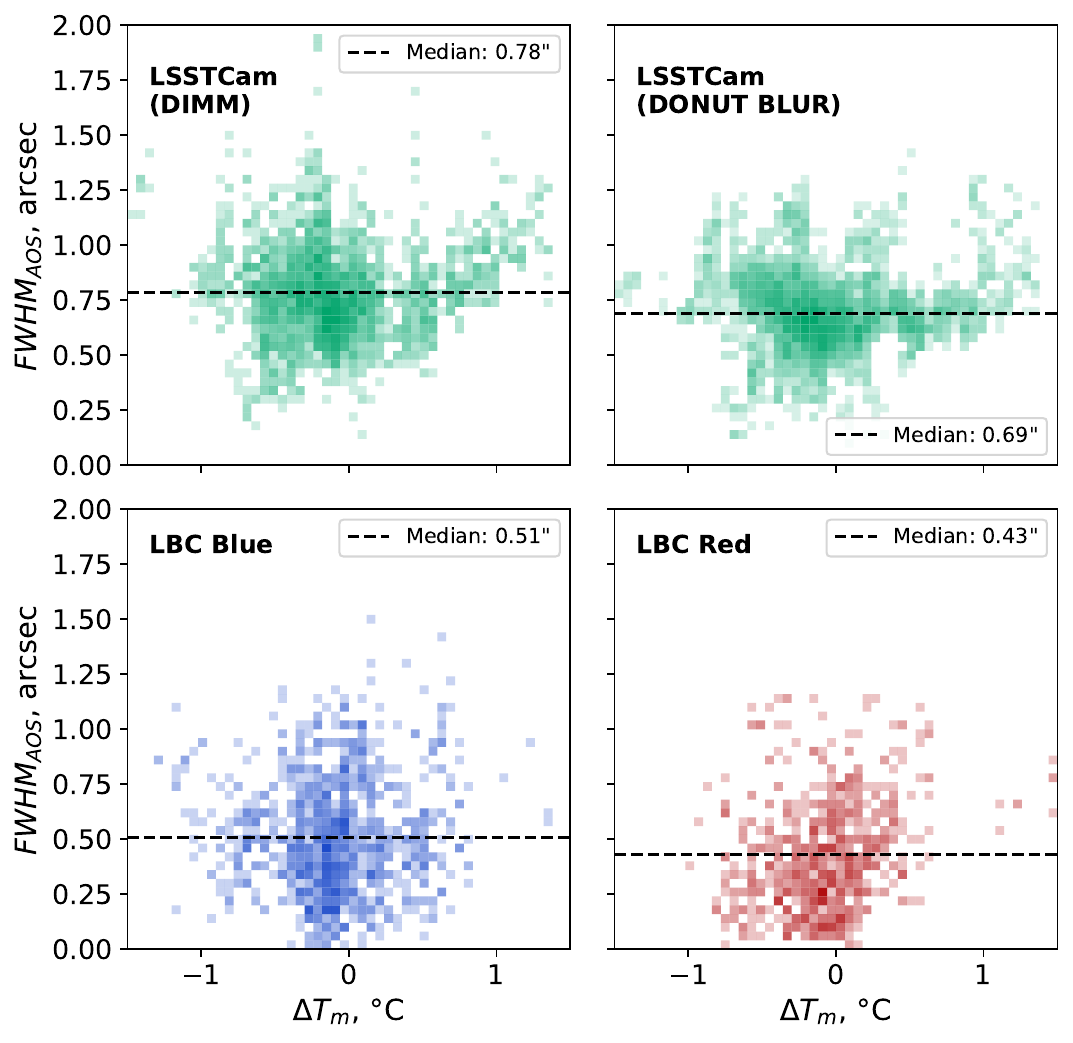}
    \caption{Relation between the $FWHM_{AOS}$ and the temperature difference between the primary mirror and the surrounding air, for LSSTCam - DIMM (a), LSSTCam - DB (b), LBC-Blue (c), and LBC-Red (d) respectively.}
    \label{fig:fwhm_aos_vs_delta_tm}
\end{figure}

\begin{figure}[htbp]
    \centering
    \includegraphics[width=0.75\textwidth]{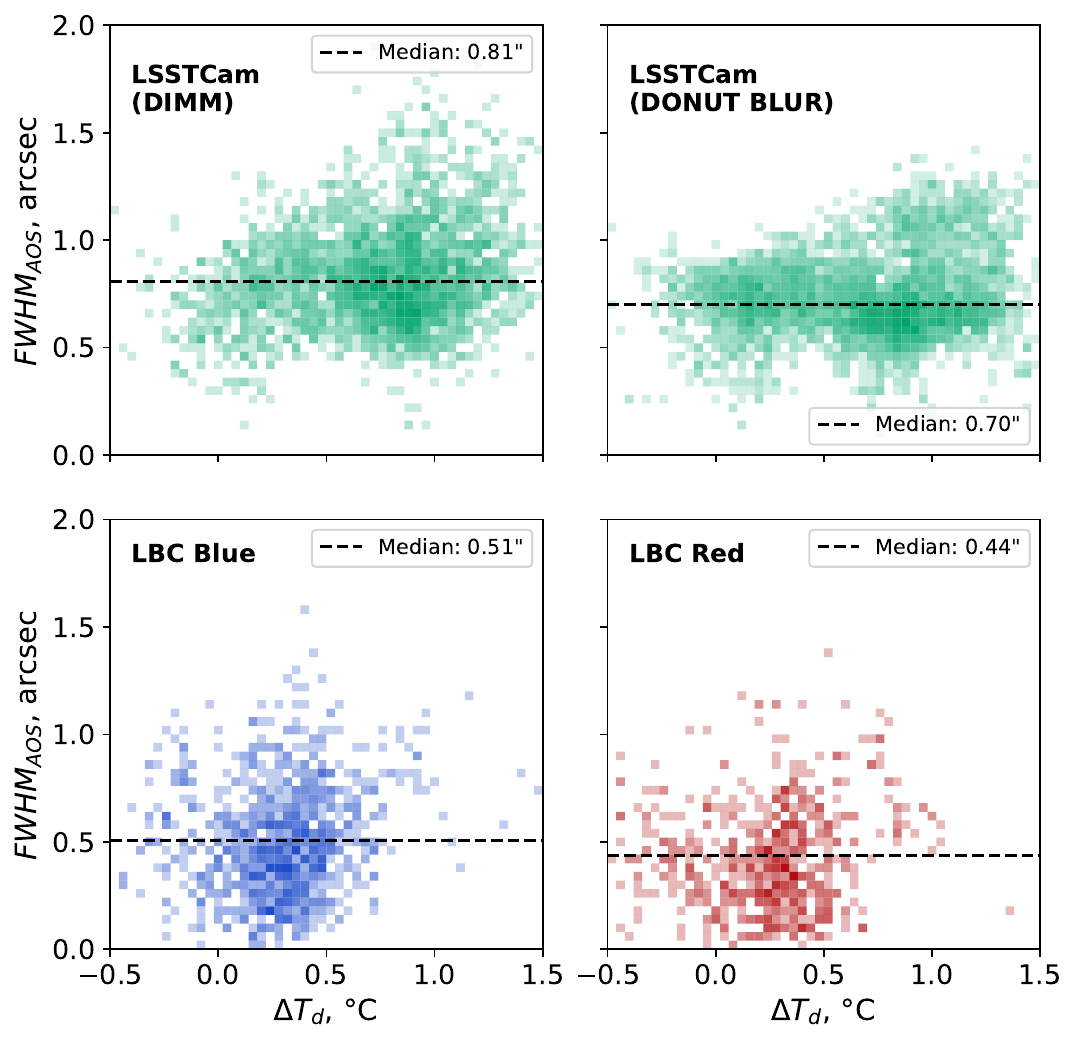}
    \caption{Relation between the $FWHM_{AOS}$ and the temperature difference between the in-dome air and outside, for LSSTCam - DIMM (top-left), LSSTCam - DB (top-right), LBC-Blue (bottom-left), and LBC-Red (bottom-right) respectively.}
    \label{fig:fwhm_aos_vs_delta_td}
\end{figure}

The temperature gradients, especially between the primary mirror and the surrounding air, and the in-dome and the outside temperature can introduce a significant amount of aberrations -- such as Spherical Aberration in the case of mirror seeing\cite{racineMIRRORDOMENATURAL1991} -- as well as create additional turbulence that increase the seeing contribution (i.e. the dome seeing). 
The primary solutions to mitigate these effects consist of efficient thermalization systems for the primary mirror and dome structures designed for a good ventilation, usually by means of louvers. In addition, a set of open-loop corrections -- like for the gravity load compensations -- can be built and applied to correct the mirrors figures depending on the measured temperature gradients. Eventually, the residual aberrations can be compensated with the AOS closed-loop corrections. 
Figure \ref{fig:fwhm_aos_vs_delta_tm} shows the relations between the $FWHM_{AOS}$ and the temperature difference between the primary mirror and the in-dome air ($\Delta T_m$), while Figure \ref{fig:fwhm_aos_vs_delta_td} shows the dependency on the temperature difference between inside and outside the dome ($\Delta T_d$).
For both LBT and Rubin the majority of the thermal gradients is within the 1°C, where the mirror seeing effect is not significant\cite{racineMIRRORDOMENATURAL1991}. The beginning of an increased trend is visible for $\Delta T_m > 1.0$ for Rubin, more significant for the DIMM case w.r.t. the DB one. 
Regarding the inside-to-outside thermal gradient, LBT has the majority of the data points within the 0.0-0.5°C range while Rubin sets between 0.5-1.0 °C. This could be a symptom of the not-yet fully operative ventilation system of Rubin.  Moreover, a significant amount of large $FWHM_{AOS}$ is present for $\Delta T_d > 0.5$ °C, supporting the idea that the dome seeing could be a significant effect of the $FWHM_{AOS}$ budget.

\subsection{Time dependency}\label{subsec:delta_time_dep}
\begin{figure}[htbp]
    \centering
    \includegraphics[width=0.7\textwidth]{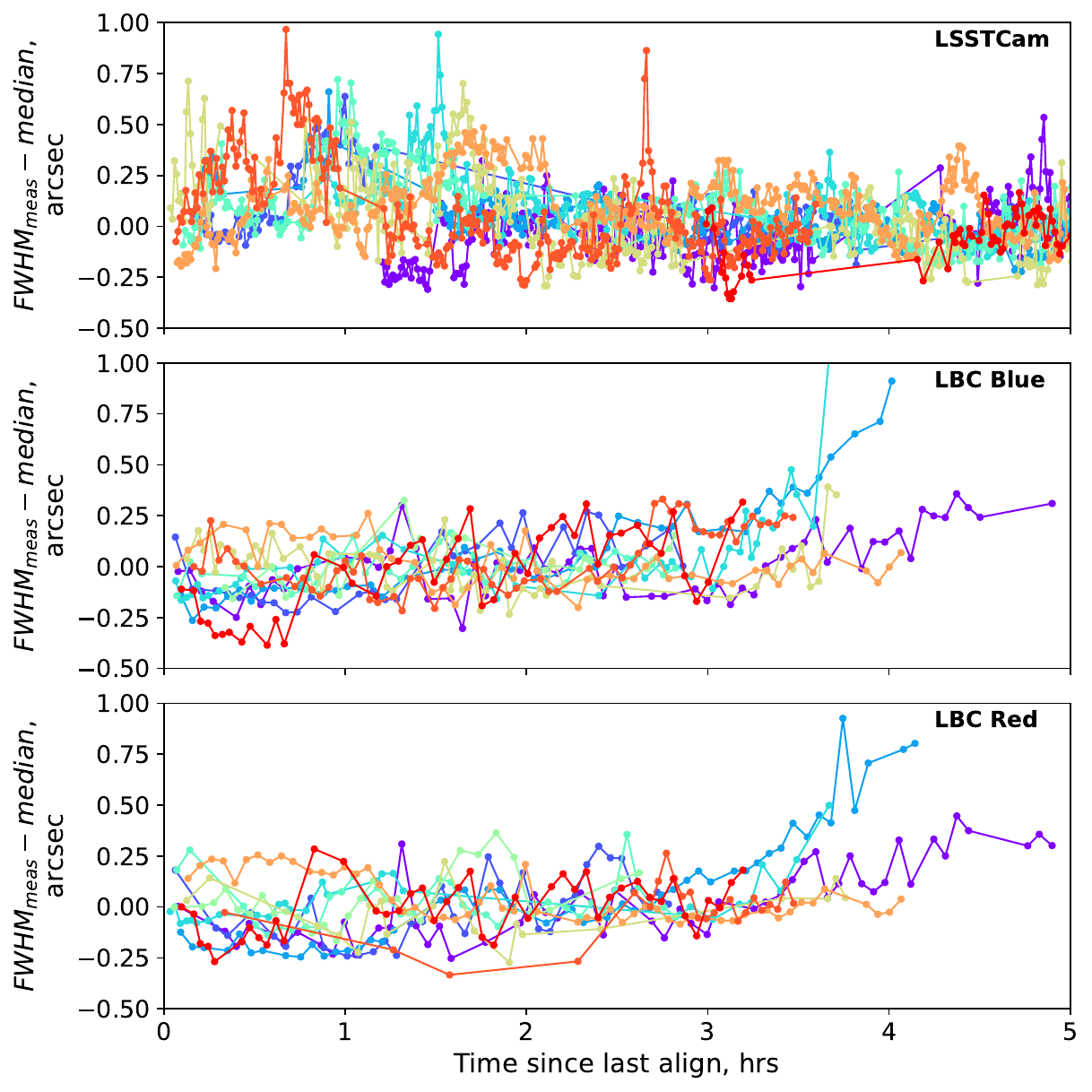}
    \caption{Relation between the $FWHM_{meas}$ (normalized for the median value of each subset) and the time elapsed from the last alignment procedure, for LSSTCam (top), LBC-Blue (central) and LBC-Red (bottom), respectively.}
    \label{fig:fwhm_meas_vs_time}
\end{figure}

Finally, we evaluated the temporal stability of the AOS from the last active alignment procedure. Usually, the AOS send corrections in between the images, maintaining the good optical state for a long time, even many hours. A re-collimation is sometimes necessary to offload the active components, or when a big change in the telescope state occurs, such as a long slew. The Rubin's telescope is a particular case: being a survey telescope, it randomly walks across the sky, and it requires the telescope to settle within a few seconds, regardless of the previous position\cite{thomasRubinObservatorySimonyi2023b}. LBT is a traditional pointing telescope, tracking for a long time on single objects across the night. Before the advent of the TMS, a re-collimation through the FPIA was made every 0.5 h, but the TMS pushed this interval towards the 1-3 hours\cite{rakichCommissioningLaserMetrology2022}, being able to correct in real-time the rigid body misalignment between the optics.
Figure \ref{fig:fwhm_meas_vs_time} shows the trend of the on-axis $FWHM_{meas}$ against the time elapsed from the last telescope alignment. The values are normalized for the median value of each subset, to enhance any possible trend. We restricted the dataset to 10 nights to keep the figure more clean. 
The stability of the Rubin AOS is remarkably good, showing no significant increase of $FWHM_{meas}$ with time. On the other side, LBT shows the effect of the TMS, maintaining the a good optical quality up to 1.5-2 hours, but after this time the collimation starts to degrade. 


\section{CONCLUSIONS}\label{sec:conclusions}
This work demonstrates the solidity of both AOS, and highlights their performance differences.
While the metrology of LBT keeps the telescope alignment regardless of the seeing condition and telescope elevation (within certain limits), the capability of Rubin's AOS closed loop to maintain the telescope's optimal state for the entire night is certain, regardless of the random walk of the telescope across the sky.

Moreover, this work also reflects the different evolutionary stages of LBT and Rubin. On a short-to-medium-term timescale (from a few months to a year), this comparative study is expected to evolve significantly. In particular for the performance of the Rubin Observatory. Currently, it is navigating the final, complex stage of its performance optimization. The completion of the ventilation system of the dome, the refinement of the CWFS algorithm, and the full empirical calibration of the gravity and temperature-dependent LUTs will significantly improve the telescope's response to environmental perturbations. Therefore, once the Rubin achieves its complete operational maturity, the current analysis would be worth re-executing.

Another pivotal aspect of finalizing this comparative landscape involves the full, routine integration of the alignment using the LT for the Rubin telescope. Currently, the initial nighttime alignment at the Rubin Observatory is not performed using the LT systematically, due to a pending but fundamental optimization of this system that is now under testing. In the interim, the standard operational procedure relies on an empirical workaround: restoring the telescope's opto-mechanical structure to the best-performing alignment configuration recorded during the previous night.
Once the LT is adopted as the initial baseline alignment, the core premise of this architectural comparison—evaluating the LBT's paradigm (initial optical wavefront sensing maintained by continuous laser metrology) directly against Rubin's intended paradigm (initial laser metrology maintained by continuous optical wavefront sensing)—can be conclusively evaluated.
Furthermore, this operational shift presents a unique opportunity for future research. Conducting a dedicated analysis of Rubin’s AOS performance, specifically comparing the current "restoration" alignment against the future pure LT alignment, will be of great interest.
On the LBT side, an new implementation of the TMS between the primary mirror and the Adaptive Secondary Mirror (ASM) is under commissioning. This new configuration will unlock the possibility to have at the same time the TMS telemetry and the wavefront sensing data from the Wavefront Sensors (WFSs) of the Gregorian instruments. This configuration will provide a unique case of study to understand the complementarity of these two worlds.

\acknowledgments 
This material is based upon work supported in part by the National Science Foundation through Cooperative Agreements AST-1258333 and AST-2241526 and Cooperative Support Agreements AST-1202910 and 2211468 managed by the Association of Universities for Research in Astronomy (AURA), and the Department of Energy under Contract No. DE-AC02-76SF00515 with the SLAC National Accelerator Laboratory managed by Stanford University. Additional Rubin Observatory funding comes from private donations, grants to universities, and in-kind support from LSST-DA Institutional Members.\\
The authors acknowledge the contribution of the USC A of the INAF Scientific Directorate to the program "Italian participation in the Rubin LSST project".\\
This work was supported by INAF funds for the MORFEO project ELT MORFEO INAF 1.05.03.19.01.
PNRR funds : Avviso N° 3264 28-12-2021 PNRR M4C2 Riferimento PNRR IR0000034 STILES Investimento 3.1 CUP C33C22000640006.
I thank Dr. Emiliano Diolaiti for its useful insights on the LBC cameras and to Andrew Rakich for pioneering the GMT and LBT TMS.

\bibliography{report} 
\bibliographystyle{spiebib} 

\end{document}